\documentclass[aps,prl,twocolumn,showpacs,floatfix]{revtex4-1}

\usepackage{times}

\usepackage[english]{babel}
\usepackage{amsmath,amssymb,bbm}

\newcommand{\mathd}{\mathrm{d}}
\newcommand{\mathe}{\mathrm{e}}
\newcommand{\tmem}[1]{{\em #1\/}}
\newcommand{\tmop}[1]{\ensuremath{\operatorname{#1}}}
\newcommand{\tmtextbf}[1]{{\bfseries{#1}}}

\begin{document}
\title{Generalized master equations leading to completely positive
dynamics}

\author{Bassano Vacchini}
\affiliation{Dipartimento di Fisica, Universit\`a degli Studi di Milano,
Via Celoria 16, I-20133 Milan, Italy}
\affiliation{INFN, Sezione di Milano, Via Celoria 16, I-20133 Milan,
  Italy}

\begin{abstract}
We provide a general construction of quantum generalized master equations with memory kernel leading to well defined, that is completely positive and trace preserving, time evolutions. The approach builds on an operator generalization of memory kernels appearing in the description of non-Markovian classical processes, and puts into evidence the non uniqueness of the relationship arising due to the typical quantum issue of operator ordering. The approach provides a physical interpretation of the structure of the kernels, and its connection with the classical viewpoint allows for a trajectory description of the dynamics. Previous apparently unrelated results are now connected in a unified framework, which further allows to phenomenologically construct a large class of non-Markovian evolutions taking as starting point collections of time dependent maps and instantaneous transformations describing the microscopic interaction dynamics.
\end{abstract}
 
\pacs{03.65.Yz, 02.50.-r, 42.50.Lc, 03.65.Ta}
\date{\today}
\maketitle

In the presence of an external environment
the time evolution of a quantum system is no more given by a
reversible unitary dynamics. For the description of such open quantum
systems one of the important issues is the determination of
equations providing a well-defined reduced dynamics
{\cite{Breuer2002}}. While for a reversible quantum evolution Stone's
theorem implies that the determination of the time evolution amounts
to the identification of the system Hamiltonian, no such general
result is available for a generic reduced dynamics. Such a result
would be of major importance also in view of phenomenological
approaches, since the very complexity of a general system environment
setting suggests that a microscopic approach starting from a
Hamiltonian description for both system and environment is often
unfeasible. Indeed, while perturbative techniques are known in order
to formally obtain the reduced dynamics of the system degrees of
freedom both in the form of integro-differential equations and of
time-local master equations {\cite{Breuer2002}},
the perturbative analysis is quite
cumbersome and in particular preservation of complete positivity (CP) is not
warranted unless all terms of the perturbation expansion are considered. The
property of CP {\cite{Nielsen2000}} ensures positivity of the time evolution
in the presence of an arbitrary ancillary system regardless of its interaction
with the system of interest. Given a factorized initial system-environment
state and a unitary interaction between system and environment the reduced
time evolution has to be CP {\cite{Holevo2001}}. It is therefore natural to
ask phenomenological evolution equations to preserve this property.
A key characterization has been given for the case in which the
evolution maps combine as $\Phi (t +
s) = \Phi (t) \Phi (s)$ for positive times only, corresponding to a
semigroup composition law. 
The most general
expression for a semigroup
of quantum CP transformations is given by $\Phi (t) =
\mathe^{\mathcal{L} t}$, where
the socalled Lindblad generator $\mathcal{L}$ solves the master
equation $\dot{\rho} (t) = \mathcal{L} \rho (t)$, and its structure is
fixed by
a famous theorem
{\cite{Gorini1976a-Lindblad1976a}}.
Further important results on the possible structure of time-local master
equations leading to a well-defined CP dynamics have been obtained. 
In such a case hermiticity and trace preservation already strongly constrains
the operator structure of the equation, and this has allowed to determine quite
general sufficient conditions warranting the existence of a CP reduced
dynamics {\cite{Breuer2007a-Andersson2007a-Vacchini2012a-Wissmann2015a}}.
Basically one considers a master equation whose operator structure is the same
as in the semigroup case, but coefficients and operators can now depend on
time and one looks for conditions on this time dependence warranting CP. Much
less is known in the case of generalized master equations of the form
\begin{equation}
  \frac{\mathd}{\mathd t} \rho (t) = \int^t_0 \mathd \tau \mathcal{K}
  (t - \tau) \rho (\tau),  \label{eq:k}
\end{equation}
where the operator $\mathcal{K} (t)$ is called memory kernel (MK), possibly
including at the r.h.s. a term of the form $\mathcal{I} (t) \rho (0)$,
that is a inhomogeneous contribution. In this framework even the requirement of
hermiticity and trace preservation is not easily satisfied, let alone CP.
Moreover in this case one often lacks a simple connection between the
expression of the MK and the basic microscopic physical interaction
mechanisms, at variance with the time-local case, in which coefficients can
often naturally be interpreted as rates and the so-called Lindblad operators
appearing in the structure can typically be connected e.g. with transitions among
system states. This fact further
hindered the determination of well-defined MK on the basis of physical
intuition, and indeed innocent looking or apparently physically motivated MK
actually lead to ill defined time evolutions
{\cite{Barnett2001a-Shabani2005a-Campbell2012a}}. Despite this, well-defined
generalized master equations have been
obtained, both within mathematical or phenomenological approaches
{\cite{Daffer2004a,Budini2004a,Budini2005a,Breuer2008a,Wilkie2009a,Budini2013b,Vacchini2013a,Chruscinski2016a}}
and considering definite microscopic models
{\cite{Cresser1996a,Giovannetti2012a-Giovannetti2012b-Ciccarello2013a-Ciccarello2013b,Lorenzo2016a}}.
However a general construction both encompassing known examples and providing
hints for the determination of generalized master equations based on the
introduction of quantum maps, to be guessed phenomenologically or determined
from microscopic physical interactions, is yet not available. The determination of time
evolutions beyond a semigroup law is also relevant in order to
describe quantum memory effects
{\cite{Rivas2014a-Breuer2016a}}.
In this Letter we show how to obtain general classes of quantum
MK master equations building on the structure of classical 
MK leading to non-Markovian classical processes. In moving from the
classical to the quantum realm the correspondence between classical quantities
and quantum operators is not unique and the different viable options of
operator ordering lead to a rich structure, reflecting themselves in the different time ordering of the operators in
the solution. Indeed this subtle issue allows to understand and connect
apparently unrelated results. Coming from classical non-Markovian processes
also a trajectory viewpoint is naturally available, allowing a physical
interpretation of the operators determining the MK. Our result
further shows how much can be learnt coming to quantum mechanics from a
classical probabilistic viewpoint {\cite{Holevo2001}}. 

\paragraph*{Classical and quantum memory kernel}Let us consider a
classical system living on a denumerable set of states. Once in a
state $k$ it will remain there for a time determined by a probability
distribution $f_k (t)$, called waiting time distribution, and then jump to
another state $n$ with a probability given by the element $\pi_{nk}$ of
a given stochastic matrix. To each waiting time distribution $f_k (t)$ is
associated its survival probability $g_k (t) = 1 - \int_0^t d \tau f_k
(\tau)$, providing the probability not to leave the state up to time $t$. The
conditional transition probability of the process $T_{nm} (t)$, namely the
probability to be in state $n$ at time $t$ under the condition of starting in
$m$ at time zero, obeys the integro-differential equation
{\cite{Feller1964a-Gillespie1977a-Breuer2009a}}
\begin{equation}
  T_{nm} (t) = \delta_{nm} g_m (t) + \int_0^t d \tau \sum_k w_{nk} (t - \tau)
  T_{km} (\tau), \label{eq:start}
\end{equation}
where the function $w_{nk} (t)$ has a simple expression in Laplace transform,
namely $\hat{w}_{nk} (u) = \hat{g}_n (u) \pi_{nk} \hat{f}_k (u) / \hat{g}_k
(u)$. The case of a Markovian process is then recovered for waiting time
distributions of exponential form with rate $\lambda_k$, corresponding to
$w_{nk} (t) = e^{- \lambda_n  (t - \tau)} \pi_{nk} \lambda_k$. Considering a
process starting in a fixed state, as described in 
the Supplemental Material
{\cite{supp}} its probability vector $P_n (t)$ obeys the same generalized
master equation, which in Laplace transform reads
\begin{multline}
  u \hat{P}_n (u) - P_n (0) =
\\ \sum_m \left[ \pi_{nm}  \frac{\hat{f}_m
  (u)}{\hat{g}_m (u)} - \delta_{nm}  \left( \frac{1}{\hat{g}_m (u)} - u
  \right) \right]  \hat{P}_m (u), 
\label{eq:sp}
\end{multline}
providing a convenient starting point for a quantum generalization. Indeed
written in this way the MK is determined by quantities, such as stochastic
matrix and waiting time distribution, admitting a direct physical
interpretation. It further warrants that the solution $P_n (t)$ is at any time
a well-defined probability vector and arises from a reading of the time
evolution in term of trajectories, corresponding in particular to examples of
so-called semi-Markov processes {\cite{Cox1965}}. At variance with
{\cite{Breuer2008a}}, where the existence of such processes was a
motivation to look for quantum MK in the form of time-dependent Lindblad
generators, further pointing to conditions on the warranting of CP based on a
perturbative analysis of the solution, we will here more closely focus on the
specific form of the MK appearing in Eq.~(\ref{eq:sp}), thus in particular
keeping the connection with a trajectory viewpoint. This aspect was partially
developed in {\cite{Vacchini2013a}}, though fully missing the deep connection
with the MK of classical non-Markovian processes and inadvertently using a
particular operator ordering. 

In quantum mechanics probability vectors are replaced by statistical
operators and in order to obtain suitable MK for Eq.~(\ref{eq:k})
one can start from~(\ref{eq:sp}) replacing the different $\mathbb{C}$-number quantities by
operator-valued ones according to
\begin{equation}
  \label{eq:op}
 \widehat{\mathcal{K}} (u)= \mathcal{O} [\pi (\hat{f} (u) / \hat{g} (u))] -\mathcal{O} [\widehat{g \mathcal{G}} (u)^{- 1}
  - u] 
\end{equation}
where $\mathcal{O} [\cdot]$ denotes an operator replacement rule also keeping
into account the issue of operator ordering. This ordering will determine the
distribution of the action in time of the different non commuting operators.
Note that, while the dynamics will be defined in terms of time dependent
operators, it is convenient for the sake of simplicity to introduce the
replacement rule in Laplace transform. The quantum counterpart of the
stochastic matrix $\pi$ is an arbitrary CP trace preserving transformation
$\mathcal{E}$, while the waiting time distribution$f (t)$ will be replaced by
$f (t) \mathcal{F} (t)$, with $\mathcal{F} (t)$ a collection of time dependent
CP trace preserving maps describing the transformation of the system between
jumps. Similarly the survival probability $g (t)$ goes over to $g (t)
\mathcal{G} (t)$, where again the maps $\mathcal{G} (t)$ are CP trace
preserving and such that $\mathcal{G} (0) = \mathbbm{1}$. We then consider the
following operator replacement rule
\begin{eqnarray}
  \mathcal{O} [\pi (\hat{f} (u) / \hat{g} (u))] & \rightarrow & \widehat{g
  \mathcal{G}} (u)^{- 1} \widehat{f \mathcal{F}} (u) \mathcal{E}, 
  \label{eq:r}
\end{eqnarray}
where $\widehat{f \mathcal{F}} (u)$ denotes the Laplace transform of $f (t)
\mathcal{F} (t)$ and similarly for $g (t) \mathcal{G} (t)$, leading to
\begin{eqnarray}
  \widehat{\mathcal{K}}_R (u) & = & \widehat{g \mathcal{G}} (u)^{- 1}
  \widehat{f \mathcal{F}} (u) \mathcal{E} - (\widehat{g \mathcal{G}} (u)^{- 1}
  - u),  \label{eq:kr}
\end{eqnarray}
where note that operator ordering only plays a role in the first term at the
r.h.s. of Eq.~(\ref{eq:op}). This operator MK immediately leads to the
expression of the time evolution map {\cite{supp}} transforming the initial
quantum state in the time evolved one
\begin{eqnarray}
  \hat{\Phi}_R (u) & = & (\mathbbm{1} - \widehat{f \mathcal{F}} (u)
  \mathcal{E})^{- 1} \widehat{g \mathcal{G}} (u),  \label{eq:fr}
\end{eqnarray}
so that in the time domain, replacing the inverse by a Neumann series and
exploiting the fact that multiplication goes over to convolution, we obtain
for $\rho (t) = \Phi_R (t) \rho (0)$
\begin{equation}
  \rho (t) = \sum_{n = 0}^{\infty} (\ast^n (f\mathcal{F}\mathcal{E}) \ast
  (g \mathcal{G})) (t) \rho (0), \label{eq:rrt}
\end{equation}
where $\ast^n$ denotes the $n$-fold convolution. This very
expression warrants CP of $\Phi_R (t)$, as composition of CP maps, while the
requirement of trace preservation, calling for a kind of balance between the
two contributions of Eq.~(\ref{eq:kr}), can be read directly from the kernel
as shown in {\cite{supp}} and leads to
\begin{eqnarray}
  {\mathd} \tmop{Tr} \left\{ g (t) \mathcal{G} (t)  \rho
  \right\} /{\tmop{dt}}& = & - \tmop{Tr} \{ M (t) \rho \},  \label{eq:tr}
\end{eqnarray}
where $\hat{M} (u) = \widehat{g \mathcal{G}} (u)^{- 1} \widehat{f \mathcal{F}}
(u) \mathcal{E} \widehat{g \mathcal{G}} (u)$. Since $\mathcal{F} (t)$ and
$\mathcal{G} (t)$ are trace preserving~(\ref{eq:tr}) takes the simple form
\begin{eqnarray}
  {\mathd} g (t)/{\tmop{dt}} & = & - f (t),  \label{eq:fg}
\end{eqnarray}
namely just the basic relation between an arbitrary waiting time
distribution $f (t)$ and its survival probability $g (t)$. We have thus
obtained in a straightforward way a class of MK ensuring CP and trace
preservation of the associated time evolution, both non trivial requirements
in the case of integro-differential equations. It immediately appears that due
to the non commutativity of operators in quantum mechanics besides~(\ref{eq:r}), for the very same collection of time dependent maps, one can
also consider a different operator replacement
\begin{eqnarray}
  \mathcal{O} [\pi (\hat{f} (u) / \hat{g} (u))] & \rightarrow & \mathcal{E}
  \widehat{f \mathcal{F}} (u) \widehat{g \mathcal{G}} (u)^{- 1},  \label{eq:l}
\end{eqnarray}
identifying a different kernel $\mathcal{K}_L (t)$ and an evolution map
$\Phi_L (t)$ leading for $\rho (t) = \Phi_L (t) \rho (0)$ to
\begin{eqnarray}
  \rho (t) & = & \sum_{n = 0}^{\infty} ((g \mathcal{G}) \ast^n (\mathcal{E} f
  \mathcal{F})) (t) \rho (0) .  \label{eq:rlt}
\end{eqnarray}
Again one immediately has CP while the trace preservation condition takes the
simpler form $
  {\mathd} \tmop{Tr} \{ g (t) \mathcal{G} (t)   \rho
  \}/{\tmop{dt}}= - \tmop{Tr} \{ \mathcal{E} f (t) \mathcal{F} (t) \rho \}$,
still satisfied thanks to trace preservation of the single
contributions and Eq.~(\ref{eq:fg}). Kernels falling within this latter
choice have been obtained in {\cite{Chruscinski2016a}}. The two MK
thus obtained, arising from different operator orderings, indeed lead
to different dynamical evolution equations.  While in the Markovian
case the structure of quantum dynamical semigroups as quantum
counterpart of classical Markov semigroups appears to be uniquely
fixed and captured by the expression of the Lindblad generator
{\cite{Gorini1976a-Lindblad1976a}}, for the quantum counterpart of
non-Markovian classical processes a greater freedom appears. In a Lindblad master equation we only
have to fix the Lindblad operators, here the presence of a MK implies that also the time
sequence in the action of the different operators is relevant. We
stress moreover that even for fixed MK Eq.~(\ref{eq:k}) can be written
in different ways. Indeed while the
expression of the MK in Laplace transform are quite simple, in the
time domain it is convenient to introduce also a
inhomogeneous term, so that the generalized master equations
corresponding to the kernels $\mathcal{K}_{R, L} (t)$ read {\cite{supp}}
\begin{eqnarray}
  \frac{\mathd}{\mathd t} \rho (t) & = & \! \int^t_0 \mathd \tau
  \mathcal{K}_{R, L} (t - \tau) \rho (\tau) \nonumber\\
  & = & \! \int^t_0 \mathd \tau \mathcal{W}_{R, L} (t - \tau) \rho (\tau) +
  \mathcal{I} (t) \rho (0),  \label{eq:wi}
\end{eqnarray}
where the common inhomogeneous term reads
$\mathcal{I} (t) = \mathd [g (t) \mathcal{G} (t)] / \tmop{dt}$, while the
kernels $\mathcal{W}_{R, L}$ are given by
\begin{eqnarray}
  \mathcal{W}_R (t) & = & {\mathd} [f (t) \mathcal{F} (t)]
  /{\mathd t}\, \mathcal{E}+ \delta (t) f (0) \mathcal{F} (0) \mathcal{E}  \label{eq:wr}\\
  \mathcal{W}_L (t) & = & {\mathd} [H (t)]/{\mathd t},  \label{eq:wl}
\end{eqnarray}
where $H (t)$ has Laplace transform $u \widehat{g \mathcal{G}} (u) \mathcal{E}
\widehat{f \mathcal{F}} (u) \widehat{g \mathcal{G}} (u)^{- 1}$. Note the
different complexity in the kernels, which not always allow for a direct
interpretation in terms of the relevant collections of CP maps determining the
dynamics. The difference between $\mathcal{W}_R $ and
$\mathcal{W}_L $ just arises due to non commutativity, even though this
simple connection is only transparent in the Laplace domain. While further
choices can be considered moving $\mathcal{E}$ in different positions {\cite{supp}}, the one considered here is suggested by the
trajectory expansion~(\ref{eq:frt}).

An interesting case arises assuming as an Ansatz that the dynamics between the
microscopic interaction events, described by the map $\mathcal{E}$, is given
by a quantum dynamical semigroup with generator $\mathcal{L}$ in Lindblad
form. In this case the only relevant ordering depends on the positioning of
$\mathcal{E}$ with respect to the functions of $\mathcal{L}$. We can thus
consider the replacement
\begin{eqnarray}
  \mathcal{O} [\pi (\hat{f} (u) / \hat{g} (u))] & = &
  \frac{\hat{f} (u - \mathcal{L})}{\hat{g} (u - \mathcal{L})} \mathcal{E}, 
  \label{eq:olr}
\end{eqnarray}
leading to the master equation {\cite{supp}}
\begin{eqnarray}
  \frac{\mathd}{\tmop{dt}} \rho (t) & = & \mathcal{L} \rho (t) + \int^t_0
  \mathd \tau \mathe^{\mathcal{L}  (t - \tau)} k (t - \tau) \mathcal{M} \rho
  (\tau),  \label{eq:lm}
\end{eqnarray}
where the function $k (t)$ is given in Laplace transform
by $\hat{k} (u) = \hat{f} (u) / \hat{g} (u)$, in analogy with the classical
MK in~(\ref{eq:sp}), and $\mathcal{M} = (\mathcal{E} - \mathbbm{1})$ is
itself a generator in Lindblad form. The alternative choice of operator
ordering in (\ref{eq:olr}) leads to a similar
equation where the position of $\mathcal{L}$ and
$\mathcal{M}$ is exchanged. 

\paragraph*{Trajectory description and physical examples}We now want to
connect the obtained results more closely to a description of the dynamics in
terms of trajectories, further pointing to physical realizations. Let us first observe that the time evolution maps $\Phi_R (t)$ and
$\Phi_L (t)$ admit the representations
\begin{eqnarray}
  \Phi_R (t) & \!=\! & p^0_R (t) \mathcal{G} (t) + \sum_{n = 1}^{\infty} \int^t_0
  \mathd t_n \ldots \int^{t_2}_0 \mathd t_1  \label{eq:frt}\\
  &  & \times p^n_R (t ; t_n, \ldots, t_1) \mathcal{F} (t - t_n) \mathcal{E}
  \ldots \mathcal{F} (t_2 - t_1) \mathcal{E} \mathcal{G} (t_1)
       \nonumber
\\
  \Phi_L (t) & \!=\! & p^0_L (t) \mathcal{G} (t) + \sum_{n = 1}^{\infty} \int^t_0
  \mathd t_n \ldots \int^{t_2}_0 \mathd t_1  \label{eq:flt}\\
  &  & \times p^n_L (t ; t_n, \ldots, t_1) \mathcal{G} (t - t_n) 
  \ldots \mathcal{E} \mathcal{F} (t_2 - t_1) \mathcal{E} \mathcal{F} (t_1) \nonumber
\end{eqnarray}
with $p^n_{R, L} (t ; t_n, \ldots, t_1)$ the exclusive probability densities
for jumps corresponding to the action of $\mathcal{E}$ at times $t_1, \ldots,
t_n$ within the time interval from 0 to $t$. They are given by
\begin{eqnarray}
  p^n_R (t ; t_n, \ldots, t_1) & = & f (t_{} - t_n) \ldots f (t_2 - t_1) g
  (t_1)  \label{eq:pr}\\
  p^n_L (t ; t_n, \ldots, t_1) & = & g (t_{} - t_n) \ldots f (t_2 - t_1) f
  (t_1)  \label{eq:pl}
\end{eqnarray}
where the different time arguments become relevant in the
integrals~(\ref{eq:frt}) and (\ref{eq:flt}) due to connection with the
operator action. This fact embodies the further freedom available in
this situation with respect to the Markovian case. In particular~(\ref{eq:pr}) is the standard expression considered in a renewal
process describing events randomly taking place after a time interval
determined by the distribution $f (t)$. As discussed in
{\cite{Budini2013a,Vacchini2014a}} and detailed in {\cite{supp}}
Eqs.~(\ref{eq:frt}) and (\ref{eq:flt}) provide a trajectory
description of the dynamics at the level of the statistical operator
in that they express the solution of the master equation as a sum of
contributions corresponding to statistical operators determined by the
number and the time of jumps, weighted according to the probability
densities~(\ref{eq:pr}) and (\ref{eq:pl}). Each contribution is
characterized by the repeated action of the map $\mathcal{E}$ at the
given times, together with the application of the maps
$\mathcal{F} (t)$ and $\mathcal{G} (t)$ in the intermediate time
evolution, in analogy to what happens in the standard Markovian case
{\cite{Holevo2001}}.

It turns out that Eq.~(\ref{eq:frt}) corresponding to the
kernel~(\ref{eq:kr}) includes and generalizes {\cite{Vacchini2013a}},
allowing for possibly distinct collections $\mathcal{F} (t)$ and
$\mathcal{G} (t)$. It thus provides the theoretical framework
encompassing quantum collisional models
{\cite{Giovannetti2012a-Giovannetti2012b-Ciccarello2013a-Ciccarello2013b}},
including a most recently introduced generalization
{\cite{Lorenzo2016a}}, where the time evolution of the system in the
first time interval is different from those in later
ones. Conversely Eq.~(\ref{eq:flt}) describes a situation in which the
dynamics has a different characterization in the last time
interval. For a semigroup evolution among jumps, as in Eq.~(\ref{eq:lm}), 
one recovers a model of non-Markovian
dynamics first considered in a simplified case in
{\cite{Budini2004a}}. If on top of this
$\mathcal{E}$ acts as the identity, independently of the waiting time
distribution one recovers a semigroup dynamics.  Keeping a non
trivial $\mathcal{E}$ and $f (t)$ but assuming the system does not
appreciably change between jumps one obtains models of so-called
continuous time quantum random walk
{\cite{Montroll1965a,Esposito2008a}}.

The situation described by Eq.~(\ref{eq:flt}) instead, arising in presence of
the kernel $\mathcal{K}_L (t)$ determined by Eq.~(\ref{eq:l}), for the case of
an intermediate semigroup time evolution encompasses the description of
non-Markovian dynamics in the physics of the micromaser
{\cite{Cresser1992a,Herzog1995a,Cresser1996a}}. The micromaser or one-atom
maser provides one of the most fundamental systems to study light-matter
interaction {\cite{Raithel1994a-Englert2002b}}. In this system single
two-level atoms are sent through a resonant high quality single-mode microwave
cavity. The interaction between the single atoms and the cavity mode is
described by a Jaynes-Cummings Hamiltonian, and takes place for the time the
atom takes to cross the cavity, assumed to be constant. By taking the trace
with respect to the atom degrees of freedom of this unitary interaction one
obtains a CP trace preserving transformation $\mathcal{E}$.
In between the arrival of subsequent atoms the cavity mode dynamics is well
described by a semigroup evolution with a standard Lindblad generator
$\mathcal{L}$ giving the Markovian decay of the cavity field. The further
information necessary in order to determine the dynamics is the distribution
in time of the atoms flying through the cavity. For the case of a Poissonian
distribution of time of arrivals of atoms, the dynamics of the field can be
described by a Markovian master equation, as can be seen considering an
exponential waiting time distribution and assuming the semigroup assignment
$\mathcal{F} (t) = \mathcal{G} (t) = \mathe^{\mathcal{L} t}$ in
Eq.~(\ref{eq:flt}). Different distributions, allowing for non-Markovian
effects, call for a more general treatment and lead to MK master equations. Note that Eq.~(\ref{eq:flt}) can actually encompass
more general situations with respect to an intermediate semigroup evolution.
Our approach thus recovers on the one side quantum collisional models, showing
that they can be generalized to include general waiting time distributions
still leading to closed evolution equations, on
the other side a dynamics like the one of the micromaser, pointing to the fact
that it can be extended to consider situations in which the intermediate time
evolution is not necessarily of semigroup type. In particular it shows a
common path to describe the two phenomena.

We stress that the obtained results are not restricted to the case of
a finite dimensional Hilbert space, indeed while quantum collisional
models have been realized up to now considering qubit systems, in the
case of the micromaser one is actually interested in how the field
dynamics is affected by the atoms passing through the cavity. While
the expansion of the time evolution in terms of trajectories as in
Eq.~(\ref{eq:flt}) is easily obtained in this approach from the
MK~(\ref{eq:kr}), more general situations can be obtained starting
directly from Eq.~(\ref{eq:pl}) and considering rather than an
ordinary a so-called modified renewal process, in which the first
waiting time is different from the remaining ones and described by a
distribution $f_1 (t)$. In this case the evolution map is given by
\begin{eqnarray}
  \Phi (t)&& = g_1 (t) \mathe^{\mathcal{L} t} + \sum_{n = 1}^{\infty} \int^t_0
  \mathd t_n \ldots \int^{t_2}_0 \mathd t_1 \times\\
  && g (t - t_n) \mathe^{\mathcal{L}  (t - t_n)} \ldots \mathcal{E} f (t_2 - t_1)
  \mathe^{\mathcal{L}  (t_2 - t_1)}  \mathcal{E} f_1 (t_1) \mathe^{\mathcal{L}
  t_1},\nonumber \label{eq:delay}
\end{eqnarray}
with Lindblad generator $\mathcal{L}$ and jump map $\mathcal{E}$ as described
above. As detailed in {\cite{supp}} one still
obtains a closed evolution equation in integro-differential form as in
Eq.~(\ref{eq:k}) with kernel
\begin{displaymath}
  \widehat{\mathcal{K}} (u) = \mathcal{L} + [1 - \mathcal{M} (\hat{S} (u -
  \mathcal{L}) - \hat{S}_1 (u - \mathcal{L}))]^{- 1}  \mathcal{M}
  \hat{k}_1 (u - \mathcal{L}) . 
\end{displaymath}
Here $\mathcal{M}$ is defined as in~(\ref{eq:lm}), the classical kernel
$\hat{k}_1 (u) = \hat{f}_1 (u) / \hat{g} (u)$ appears operator-valued
due to the dependence on $\mathcal{L}$, while $S (t)$ is the so-called renewal
density or sprinkling distribution {\cite{Bardou2001}}, giving the
probability for a jump to occur at a given time, defined for an ordinary process
as $\hat{S} (u) = \hat{f} (u) / (1 - \hat{f} (u))$ and as $\hat{S}_1 (u) =
\hat{f}_1 (u) / (1 - \hat{f} (u))$ for a modified one, again appearing
operator-valued. For the special case of a stationary distribution of jumps
one has the constraint $f_1 (t) = g (t) / \langle \tau \rangle$, with $\langle
\tau \rangle$ the mean waiting time associated to the reference distribution
$f (t)$, leading to the model obtained in a much less straightforward way in
{\cite{Cresser1996a}}. Also here all terms appearing in the MK
have a direct meaning as physical transformation maps or
quantities related to the renewal process giving the time distribution of the jumps describing microscopic interaction events. Furthermore
despite the complicated expression of the MK both trace
preservation and CP is granted from the analysis of the ensuing dynamics in
terms of trajectories, for arbitrary waiting time distributions $f (t)$,
$f_1 (t)$ and Lindblad generator $\mathcal{L}$.

We have provided a simple general construction of quantum MK leading to well-defined reduced dynamics. The result builds on an analogy with classical non-Markovian
processes, thus allowing for a direct physical interpretation of the different
contributions appearing in the MK and for a connection to a trajectory
description of the dynamics. The interpretation of the
different kernels is best understood in Laplace domain, and can be read in the
time domain by suitably rewriting the integro-differential equation and
introducing a inhomogeneous contribution. The approach provides a general way
to build MK, complying with both trace preservation and CP, on the basis of
microscopic physical information encoded in the collection of time dependent
maps describing the time evolution in between jumps, the channel providing
the instantaneous transformation, the random distribution in time of these
transformations and the related time ordering of these maps. 
As in standard quantum mechanics, an operator replacement
rule has to be introduced, leading, at variance with the Markovian case, to a
variety of quantum stochastic dynamics corresponding to a given non-Markovian
classical one. One thus obtains a large class of non-Markovian quantum 
dynamics  including a wide range of previous results as special cases.

\section*{Acknowledgments}
The work was supported by the EU
QuProCS Project (Grant Agreement 641277) and by UniMI
H2020 Transition Grant. Motivating
discussions with M. Palma, F. Ciccarello and S. Lorenzo as well as
past correspondence with J. Cresser are also gratefully acknowledged.

\section*{Supplemental Material}

In this Supplemental Material we provide technical details on the derivation
of equations discussed in the main text of the paper.

\subsubsection*{Derivation of Eq.~(\ref{eq:sp})}

We now show how to obtain Eq.~(\ref{eq:sp}) starting from Eq.~(\ref{eq:start})
for the case of a generic waiting time distribution. In Laplace transform
Eq.~(\ref{eq:start}) takes the form
\begin{eqnarray*}
  \hat{T}_{nm} (u) & = & \delta_{nm}  \hat{g}_m (u) + \sum_k \hat{g}_n (u)
  \pi_{nk}  \frac{\hat{f}_k (u)}{\hat{g}_k (u)}  \hat{T}_{km} (u),
  \label{eq:s1}
\end{eqnarray*}
so that considering the initial condition $T_{nm} (0) = \delta_{nm}$ and
denoting by $P_n (t)$ the probability to be in state $n$ at time $t$ starting
from the state $m$ at the initial time zero one can also write this equation,
dividing by $\hat{g}_n (u)$
\begin{eqnarray*}
  \frac{\hat{P}_n (u)}{\hat{g}_n (u)} & = & P_n (0) + \sum_k \pi_{nk} 
  \frac{\hat{f}_k (u)}{\hat{g}_k (u)}  \hat{P}_k (u), \label{eq:s2}
\end{eqnarray*}
so that finally adding and subtracting terms we obtain Eq.~(\ref{eq:sp}).

\subsubsection*{Generalized master equation and trace preservation}

In order to obtain the time evolution map Eq.~(\ref{eq:fr}), once given the
operator replacement rule Eq.~(\ref{eq:r}) and the MK Eq.~(\ref{eq:kr}), let us
rewrite Eq.~(\ref{eq:k}) in Laplace transform, thus obtaining, recalling that
$\rho (t) = \Phi_R (t) \rho (0)$
\begin{eqnarray}
  u \hat{\Phi}_R (u) - \mathbbm{1} & = & \widehat{\mathcal{K}}_R (u)
  \hat{\Phi}_R (u) \label{eq:s3} \nonumber
\end{eqnarray}
and therefore
\begin{eqnarray}
  \hat{\Phi}_R (u) & = & (u - \widehat{\mathcal{K}}_R (u))^{- 1} \label{eq:s4}
  \nonumber\\
  & = & (\mathbbm{1} - \widehat{f \mathcal{F}} (u) \mathcal{E})^{- 1}
  \widehat{g \mathcal{G}} (u) \nonumber
\end{eqnarray}
thanks to Eq.~(\ref{eq:kr}). The trace preservation condition can be read from
Eq.~(\ref{eq:k}), implying that the MK has to be trace annihilating, so that
starting from Eq.~(\ref{eq:kr}) we have for any $\rho$
\begin{eqnarray*}
  0 & = & \tmop{Tr} \widehat{\mathcal{K}}_R (u) \widehat{g \mathcal{G}} (u)
  \rho  \label{eq:s5}\\
  & = & \tmop{uTr} \widehat{g \mathcal{G}} (u) \rho - 1 + \tmop{Tr}
  \widehat{g \mathcal{G}} (u)^{- 1} \widehat{f \mathcal{F}} (u) \mathcal{E}
  \widehat{g \mathcal{G}} (u) \rho, \nonumber
\end{eqnarray*}
leading to Eq.~(\ref{eq:tr}) once we define as $M (t)$ the function admitting
as Laplace transform the expression $\hat{M} (u) = \widehat{g \mathcal{G}}
(u)^{- 1} \widehat{f \mathcal{F}} (u) \mathcal{E} \widehat{g \mathcal{G}} (u)
.$ A similar calculation starting from the expression of
$\widehat{\mathcal{K}}_L (u)$ which is explicitly given by
\begin{eqnarray*}
  \widehat{\mathcal{K}}_L (u) & = & \mathcal{E} \widehat{f \mathcal{F}} (u)
  \widehat{g \mathcal{G}} (u)^{- 1} - (\widehat{g \mathcal{G}} (u)^{- 1} - u)
  \nonumber
\end{eqnarray*}
leads to
\begin{eqnarray*}
  0 & = & \tmop{Tr} \widehat{\mathcal{K}}_L (u) \widehat{g \mathcal{G}} (u)
  \rho  \label{eq:s5a}\\
  & = & u \tmop{Tr} \widehat{g \mathcal{G}} (u) \rho - 1 + \tmop{Tr}
  \mathcal{E} \widehat{f \mathcal{F}} (u) \rho, \nonumber
\end{eqnarray*}
and therefore 
the condition
 \begin{eqnarray*}
  \frac{\mathd}{\tmop{dt}} \tmop{Tr} \left\{ g (t) \mathcal{G} (t)  \! \rho
  \right\} & = & - \tmop{Tr} \{ \mathcal{E} f (t) \mathcal{F} (t) \rho \}.
\end{eqnarray*}
Let us note that also different operator
orderings with respect to Eq.~(\ref{eq:r}) and Eq.~(\ref{eq:l}), such as
\begin{eqnarray*}
  \mathcal{O} [\pi (\hat{f} (u) / \hat{g} (u))] & \rightarrow & \widehat{g
  \mathcal{G}} (u)^{- 1} \mathcal{E} \widehat{f \mathcal{F}} (u) \label{eq:sr}
  \nonumber
\end{eqnarray*}
and
\begin{eqnarray*}
  \mathcal{O} [\pi (\hat{f} (u) / \hat{g} (u))] & \rightarrow & \widehat{f
  \mathcal{F}} (u) \mathcal{E} \widehat{g \mathcal{G}} (u)^{- 1},
  \label{eq:sl} \nonumber
\end{eqnarray*}
still leading to well-defined MK and CP trace preserving
transformations. This can be seen noting the fact that given a collection of
time dependent CP trace preserving maps $\mathcal{F} (t)$,
also the maps $\mathcal{F}' (t) = \mathcal{E} \mathcal{F} (t)$ provide a
collection with the same property, so that one can use the previous
construction with the replacements $\mathcal{F} (t) \rightarrow \mathcal{F}'
(t)$ and $\mathcal{E} \rightarrow \mathbbm{1}$, and similarly considering the
maps $\mathcal{F}'' (t) = \mathcal{F} (t) \mathcal{E}$. The operator
replacement rules dealt with in detail in the main text have been chosen due to
their direct connection with physical implementations already considered in
the literature such as quantum collisional models and the field dynamics in a
micromaser.

\subsubsection*{Equivalent expressions of the generalized master equation}

The MK master equation Eq.~(\ref{eq:k}) can be written in an equivalent way by
admitting besides a convolution kernel also a inhomogeneous contribution. To
this aim one observes that while Eq.~(\ref{eq:k}) can be written in Laplace
transform as
\begin{eqnarray*}
  \widehat{\mathcal{K}} (u) & = & u - \hat{\Phi} (u)^{- 1}, \label{eq:s6}
  \nonumber
\end{eqnarray*}
once fixed a inhomogeneous term $\mathcal{I} (t)$ the time evolution map in
Laplace transform $\hat{\Phi} (u)$ determines according to
\begin{eqnarray*}
  \widehat{\mathcal{W}} (u) & = & u - (\mathbbm{1} + \widehat{\mathcal{I}}
  (u)) \hat{\Phi} (u)^{- 1} \label{eq:s7} \nonumber
\end{eqnarray*}
a new kernel $\mathcal{W} (t)$ which leads to the master equation
\begin{eqnarray*}
  \frac{\mathd}{\tmop{dt}} \rho (t) & = & \int^t_0 \mathd \tau \mathcal{W} (t
  - \tau) \rho (\tau) + \mathcal{I} (t) \rho (0) . \label{eq:s8} \nonumber
\end{eqnarray*}

\subsubsection*{Generalized master equation for the case of intermediate
semigroup dynamics}

For the case in which the collections of time dependent maps $\mathcal{F} (t)$
and $\mathcal{G} (t)$ are given by the quantum dynamical semigroup
$\mathe^{\mathcal{L} t}$, one can proceed as in {\cite{Vacchini2013a}}.
Exploiting the properties of the Laplace transform one has $\widehat{h
\mathcal{F}} (u) = \widehat{h \mathcal{}} (u - \mathcal{L})$ for any function
of $u$, so that from Eq.~(\ref{eq:fr}) we have
\begin{eqnarray*}
  \hat{\Phi} (u) & = & (\mathbbm{1} - \hat{f} (u - \mathcal{L})
  \mathcal{E})^{- 1} \hat{g} (u - \mathcal{L}) \label{eq:s10} \nonumber
\end{eqnarray*}
and according to Eq.~(\ref{eq:s6})
\begin{eqnarray*}
  \widehat{\mathcal{K}} (u) & = & u - \frac{1}{\hat{g} (u - \mathcal{L})}
  (\mathbbm{1} - \hat{f} (u - \mathcal{L}) \mathcal{E}), \label{eq:s11}
  \nonumber
\end{eqnarray*}
so that rearranging terms
\begin{eqnarray*}
  \widehat{\mathcal{K}} (u) & = & \mathcal{L} + \frac{\hat{f} (u -
  \mathcal{L})}{\hat{g} (u - \mathcal{L})} \mathcal{E} - \left(
  \frac{1}{\hat{g} (u - \mathcal{L})} - (u - \mathcal{L}) \right) \nonumber\\
  & = & \mathcal{L} + \hat{k} (u - \mathcal{L}) (\mathcal{E} - \mathbbm{1})
  \nonumber
\end{eqnarray*}
upon defining
\begin{eqnarray*}
  \hat{k} (u) & = & \frac{\hat{f} (u)}{\hat{g} (u)}, \label{eq:k2} \nonumber
\end{eqnarray*}
and further multiplying by $\hat{\rho} (u)$ and taking the inverse Laplace
transform one obtains Eq.~(\ref{eq:lm}) where $\mathcal{M} = (\mathcal{E} -
\mathbbm{1})$. A similar procedure for the alternative operator ordering
\begin{eqnarray*}
  \mathcal{O} \left[ \pi \frac{\hat{f} (u)}{\hat{g} (u)}  \right] & = &
  \mathcal{E} \frac{\hat{f} (u - \mathcal{L})}{\hat{g} (u - \mathcal{L})}
  \label{eq:oll} \nonumber
\end{eqnarray*}
leads through analogous calculations to the generalized master equation
\begin{eqnarray*}
  \frac{\mathd}{\tmop{dt}} \rho (t) & = & \mathcal{L} \rho (t) + \int^t_0
  \mathd \tau \mathcal{M} \mathe^{\mathcal{L}  (t - \tau)} k (t - \tau) \rho
  (\tau). \label{eq:s-lm} \nonumber
\end{eqnarray*}
This equation corresponds to the class of master equations used for the non-Markovian
description of the micromaser dynamics for the case in which the considered
renewal process describing the incoming atoms is not a delayed one
{\cite{Cresser1992a}}. Note that the solution of this master equation can be
written according to Eq.~(\ref{eq:flt}) as
\begin{multline*}
   \rho (t) = g (t) \mathe^{\mathcal{L} t} \rho (0)+ \! \sum_{n = 1}^{\infty}
  \int^t_0 \mathd t_n \ldots \int^{t_2}_0 \mathd t_1 \\
   \times g (t - t_n) \mathe^{\mathcal{L} (t - t_n) } \ldots \mathcal{E}f
  (t_2 - t_1) \mathe^{\mathcal{L} (t_2 - t_1) } \mathcal{E} f (t_1)
  \mathe^{\mathcal{L} t_1}\rho (0),
\end{multline*}
corresponding to Eq.~(\ref{eq:delay}) for $f_1 \rightarrow f$.

\subsubsection*{Trajectory description of the time evolution}

We here discuss in more detail how the expressions Eqs.~(\ref{eq:frt}) and
(\ref{eq:flt}) for the solutions of the MK master equations lead to a
stochastic trajectory description of the dynamics. The solution of the master
equation Eq.~(\ref{eq:k}) with MK Eq.~(\ref{eq:kr}) according to
Eq.~(\ref{eq:frt}) can be written
\begin{eqnarray*}
  \rho (t) & = & \Phi_R (t) \rho (0) \nonumber\\
  & = & p^0_R (t) \mathcal{G} (t) \rho (0) + \sum_{n = 1}^{\infty} \int^t_0
  \mathd t_n \ldots \int^{t_2}_0 \mathd t_1 \nonumber\\
  &  &\!\!\!\!\!\!\!\! \times p^n_R (t ; t_n, \ldots, t_1) \mathcal{F} (t - t_n) \mathcal{E}
  \ldots \mathcal{F} (t_2 - t_1) \mathcal{E} \mathcal{G} (t_1) \rho (0),
  \nonumber
\end{eqnarray*}
so that the statistical operator at time $t$ is given by a sum of statistical
operators $\mathcal{F} (t - t_n) \mathcal{E} \ldots \mathcal{F} (t_2 - t_1)
\mathcal{E} \mathcal{G} (t_1) \rho (0)$ corresponding to trajectories
determined by the number and the time of jumps. Indeed each contribution
arises by the action of the map $\mathcal{G}$ for a time $t_1$, followed by
jumps described by the action of the superoperator $\mathcal{E}$ at given
times $t_1, \ldots, t_n$, taking place after intermediate time evolutions
described by the action of the map $\mathcal{F}$ over a time interval given by
the difference in time of two consecutive jumps. The probability of a given
trajectory is fixed by the expression $p^n_R (t ; t_n, \ldots, t_1)$ as given
by Eq.~(\ref{eq:pr}), normalized according to
\begin{eqnarray*}
  p^0_R (t) + \sum_{n = 1}^{\infty} \int^t_0 \mathd t_n \ldots \int^{t_2}_0
  \mathd t_1 p^n_R (t ; t_n, \ldots, t_1) & = & 1, \nonumber
\end{eqnarray*}
which together with the trace preservation property of the maps $\mathcal{E}$,
$\mathcal{F} (t)$ and $\mathcal{G} (t)$ ensures normalization of $\rho (t)$.
In the Markovian case these trajectories can be obtained as solutions of a
stochastic master equation for the statistical operator, and connected to a
measurement interpretation as discussed in detail in \
{\cite{Carmichael1993a,Barchielli2009-Barchielli1994a}}, while the possible
extension of this approach to the non-Markovian case has been discussed within
the present context in {\cite{Budini2013a,Budini2013a,Vacchini2014a}}. Of
course a similar analysis holds for the statistical operator $\rho (t) =
\Phi_L (t) \rho (0)$ given by Eq.~(\ref{eq:flt}) arising as solution of
Eq.~(\ref{eq:k}) with MK obtained using the assignment Eq.~(\ref{eq:l}).

\subsubsection*{Generalized master equation for a generic delayed renewal
process}

We now want to show that the expansion Eq.~(\ref{eq:delay}), despite not
falling in the class given by the kernels Eq.~(\ref{eq:op}), still leads to a
closed evolution equation. Also in this case it is convenient to work in terms
of Laplace transforms, so that Eq.~(\ref{eq:delay}) becomes
\begin{eqnarray*}
  \hat{\Phi} (u) & = & \widehat{g_1} (u - \mathcal{L})  \label{eq:s12}\\
  &  & + \hat{g} (u - \mathcal{L}) (\mathbbm{1} - \mathcal{E} \hat{f} (u -
  \mathcal{L}))^{- 1} \mathcal{E} \widehat{f_1} (u - \mathcal{L}) \nonumber\\
  & = & \frac{1 - \widehat{f_1} (u - \mathcal{L})}{u - \mathcal{L}}
  \nonumber\\
  &  & + \frac{1 - \hat{f} (u - \mathcal{L})}{u - \mathcal{L}} (\mathbbm{1} -
  \mathcal{E} \hat{f} (u - \mathcal{L}))^{- 1} \mathcal{E} \widehat{f_1} (u -
  \mathcal{L}) \nonumber
\end{eqnarray*}
and upon multiplication by $u - \mathcal{L}$ we come to
\begin{multline*}
   (u - \mathcal{L}) \hat{\Phi} (u) - \mathbbm{1} = - \widehat{f_1} (u -
  \mathcal{L})\\
   + (1 - \hat{f} (u - \mathcal{L})) (\mathbbm{1} - \mathcal{E} \hat{f} (u -
  \mathcal{L}))^{- 1} \mathcal{E} \widehat{f_1} (u - \mathcal{L}) .
\end{multline*}
Further noticing the identity
\begin{multline*}
  (1 - \hat{f} (u - \mathcal{L}))  (\mathbbm{1} - \mathcal{E} \hat{f} (u -
  \mathcal{L}))^{- 1}  \mathcal{E}\\
  = (\mathcal{E} - \mathbbm{1})  (\mathbbm{1} - \hat{f} (u - \mathcal{L})
  \mathcal{E})^{- 1} + \mathbbm{1} 
\end{multline*}
we obtain
\begin{eqnarray*}
  (u - \mathcal{L}) \hat{\Phi} (u) - \mathbbm{1} & = & (\mathcal{E} -
  \mathbbm{1}) (\mathbbm{1} - \hat{f} (u - \mathcal{L}) \mathcal{E})^{- 1}
  \widehat{f_1} (u - \mathcal{L}) \nonumber
\end{eqnarray*}
and consider the rewriting
\begin{multline*}
  (\mathbbm{1} - \hat{f} (u - \mathcal{L}) \mathcal{E})^{- 1}\\
  = [1 - \hat{S} (u - \mathcal{L}) (\mathcal{E} - \mathbbm{1})]^{- 1} 
  (\mathbbm{1} - \hat{f} (u - \mathcal{L}))^{- 1},
\end{multline*}
where we have introduced the so-called renewal density or sprinkling
distribution {\cite{Bardou2001}}
\begin{eqnarray*}
  \hat{S} (u) & = & \frac{\hat{f} (u)}{1 - \hat{f} (u)} . \nonumber
\end{eqnarray*}
We can now express $\hat{\Phi} (u)$ introducing everywhere the more compact
notation $\mathcal{M} = (\mathcal{E} - \mathbbm{1})$, thus obtaining
\begin{multline*}
  (u - \mathcal{L}) \hat{\Phi} (u) - \mathbbm{1}  =  [1 - \mathcal{M}
  \hat{S} (u - \mathcal{L})]^{- 1} \mathcal{M} \frac{\widehat{f_1} (u -
  \mathcal{L})}{1 - \hat{f} (u - \mathcal{L})}  \label{eq:s13}\\
   =  [1 - \mathcal{M} \hat{S} (u - \mathcal{L})]^{- 1} \mathcal{M}
  \widehat{k_1} (u - \mathcal{L}) \frac{1}{u - \mathcal{L}} \nonumber
\end{multline*}
where
\begin{eqnarray*}
  \widehat{k_1} (u) & = & \frac{\widehat{f_1} (u)}{\hat{g} (u)} . \nonumber
\end{eqnarray*}
As a last step, in order to obtain a closed evolution equation for the
statistical operator, we need to reexpress the r.h.s. of the previous expression as
the action of a map on $\hat{\Phi} (u)$. To this aim we multiply both sides
by $[1 - \mathcal{M} \hat{S} (u - \mathcal{L})]$ and suitably add and subtract
a term, coming to
\begin{multline*}
  \left[ 1 - \mathcal{M} \hat{S} (u - \mathcal{L}) + \mathcal{M} \widehat{k_1}
  (u) \frac{1}{u - \mathcal{L}} \right]  [(u - \mathcal{L}) \hat{\Phi} (u) -
  \mathbbm{1}]\\
  = \mathcal{M} \widehat{k_1} (u) \hat{\Phi} (u),
\end{multline*}
and thus finally
\begin{multline*}
   [(u - \mathcal{L}) \hat{\Phi} (u) - \mathbbm{1}]\\
   = [1 - \mathcal{M} (\hat{S} (u - \mathcal{L}) - \widehat{S_1} (u -
  \mathcal{L}))]^{- 1} \mathcal{M} \widehat{k_1} (u) \hat{\Phi} (u)
\end{multline*}
upon defining
\begin{eqnarray*}
  \widehat{S_1} (u) & = & \frac{\widehat{f_1} (u)}{1 - \hat{f} (u)} =
  \frac{\widehat{k_1} (u)}{u}, \nonumber
\end{eqnarray*}
so that one recovers the expression of the kernel given below Eq.~(\ref{eq:delay}).


\begin{thebibliography}{10}
  \bibitem{Breuer2002}H.-P. Breuer and F.~Petruccione, {\tmem{The Theory of
  Open Quantum Systems}} (Oxford University Press, Oxford, 2002)
  
  
  
  
  
  \bibitem{Nielsen2000}M.~Nielsen and I.~Chuang, {\tmem{Quantum Computation
  and Quantum Information}} (Cambridge University Press, Cambridge, 2000)
  
  \bibitem{Holevo2001}A.~S. Holevo, {\tmem{Statistical Structure of Quantum
  Theory}}, Vol. m 67 of {\tmem{Lecture Notes in Physics}} (Springer, Berlin,
  2001)
  
  \bibitem{Gorini1976a-Lindblad1976a}V.~Gorini \textit{et al.},
  J.~Math. Phys. \tmtextbf{17}, 821 (1976); G.~Lindblad, Comm. Math. Phys. \tmtextbf{48}, 119
  (1976)
  
  \bibitem{Breuer2007a-Andersson2007a-Vacchini2012a-Wissmann2015a}H.-P. Breuer,
    Phys. Rev.~A \tmtextbf{75}, 022103 (2007); E.~Andersson, J.~D. Cresser, and M.~J.~W. Hall,
  J.~Mod. Opt. \tmtextbf{54}, 1695 (2007); B.~Vacchini, J. Phys. B \tmtextbf{45}, 154007
  (2012); S.~Wi{\ss}mann, H.-P. Breuer, and B.~Vacchini,
  Phys. Rev. A \tmtextbf{92}, 042108 (2015)
  
  \bibitem{Barnett2001a-Shabani2005a-Campbell2012a}S.~M. Barnett and S.~Stenholm, Phys. Rev.~A
  \tmtextbf{64}, 033808 (2001); A.~Shabani and D.~A. Lidar, Phys. Rev.~A
  \tmtextbf{71}, 020101 (2005); S.~Campbell \textit{et al.}, Phys. Rev. A \tmtextbf{85},
  032120 (2012)
 
 \bibitem{Daffer2004a}S.~Daffer \textit{et al.}, Phys. Rev.~A \tmtextbf{70}, 010304 (2004)
  
  \bibitem{Budini2004a}A.~A. Budini, Phys. Rev.~A \tmtextbf{69}, 042107
  (2004)
  
  \bibitem{Budini2005a}A.~A. Budini, Phys. Rev.~E \tmtextbf{72}, 056106
  (2005)
  
  \bibitem{Breuer2008a}H.-P. Breuer and B.~Vacchini, Phys. Rev. Lett.
  \tmtextbf{101}, 140402 (2008)
  
  \bibitem{Wilkie2009a}J.~Wilkie and Y.~M. Wong, J.~Phys.~A: Math. Gen.
  \tmtextbf{42}, 015006 (2009)
  
  \bibitem{Budini2013b}A.~A. Budini, Phys. Rev. A \tmtextbf{88}, 012124
  (2013)
  
  \bibitem{Vacchini2013a}B.~Vacchini, Phys. Rev.~A \tmtextbf{87},
  030101(R) (2013)
  
  \bibitem{Chruscinski2016a}D.~Chruscinski and A.~Kossakowski, Phys. Rev.
  A \tmtextbf{94}, 020103(R) (2016)
  
  \bibitem{Cresser1996a}J.~D. Cresser and S.~M. Pickles, J. Opt. B: Quantum
  Semiclass. Opt. \tmtextbf{8}, 73 (1996)
  
  
  \bibitem{Giovannetti2012a-Giovannetti2012b-Ciccarello2013a-Ciccarello2013b}V.~Giovannetti and G.~M. Palma, Phys. Rev.
  Lett. \tmtextbf{108}, 040401 (2012); V.~Giovannetti and G.~M. Palma, J. Phys. B
  \tmtextbf{45}, 154003 (2012); F.~Ciccarello, G.~M. Palma, and V.~Giovannetti,
  Phys. Rev. A \tmtextbf{87}, 040103 (2013); F.~Ciccarello and V.~Giovannetti, Physica
  Scripta \tmtextbf{T153}, 014010 (2013)
  
  \bibitem{Lorenzo2016a}S.~Lorenzo, F.~Ciccarello, and G.~M. Palma, Phys.
  Rev. A \tmtextbf{93}, 052111 (2016)
  
  \bibitem{Rivas2014a-Breuer2016a}A.~Rivas, S.~F. Huelga, and M.~B. Plenio, Rep. Prog.
  Phys. \tmtextbf{77}, 094001 (2014); H.-P. Breuer, E.-M. Laine, J.~Piilo, and
  B.~Vacchini, Rev. Mod. Phys. \tmtextbf{88}, 021002 (2016)
  
  \bibitem{Feller1964a-Gillespie1977a-Breuer2009a}W.~Feller, PNAS \tmtextbf{51}, 653 (1964); D.~T. Gillespie, Phys. Lett.~A \tmtextbf{64}, 22
  (1977); H.-P. Breuer and B.~Vacchini, Phys. Rev.~E
  \tmtextbf{79}, 041147 (2009)
  
  \bibitem{supp}{\tmem{See Supplemental Material [url] which incudes Refs.~\cite{Carmichael1993a,Barchielli2009-Barchielli1994a}}}
  
  \bibitem{Cox1965}D.~R. Cox and H.~D. Miller, {\tmem{The theory of
  stochastic processes}} (John Wiley \& Sons Inc., New York, 1965)
  
  \bibitem{Budini2013a}A.~A. Budini, Phys. Rev. A \tmtextbf{88}, 032115
  (2013)
  
  \bibitem{Vacchini2014a}B.~Vacchini, Int. J. Quantum Inf. \tmtextbf{12}, 10 (2014)
  
  \bibitem{Montroll1965a}E.~W. Montroll and G.~H. Weiss, J.~Math. Phys.
  \tmtextbf{6}, 167 (1965)
  
  \bibitem{Esposito2008a}M.~Esposito and K.~Lindenberg, Phys. Rev.~E
  \tmtextbf{77}, 051119 (2008)
  
  \bibitem{Cresser1992a}J.~D. Cresser, Phys. Rev. A \tmtextbf{46}, 5913
  (1992)
  
  \bibitem{Herzog1995a}U.~Herzog, Phys. Rev. A \tmtextbf{52}, 602 (1995)
  
  \bibitem{Raithel1994a-Englert2002b}G.~Raithel \textit{et al.}, in
    {\tmem{Cavity Quantum Electrodynamics}}, edited by P.  R. Berman
    (Academic Press, San Diego, 1994), pp. 57--121; B.-G. Englert and
    G.~Morigi, {\tmem{Five lectures on dissipative master equations}},
    in {\tmem{Coherent Evolution in Noisy Environments}}, edited by
    A. Buchleitner and K. Hornberger (Springer, Berlin, 2002), Lecture
    Notes in Physics 611, pp. 55--106

    \bibitem{Bardou2001}F.~Bardou \textit{et al.}, {\tmem{L{\'e}vy
          Statistics and Laser Cooling}}
  (Cambridge University Press, 2001)

 
  \bibitem{Carmichael1993a}H.~Carmichael, {\tmem{An Open Systems
        Approach to Quantum Optics}} (Springer, Berlin, 1993)
  
  \bibitem{Barchielli2009-Barchielli1994a}A.~Barchielli and M.~Gregoratti, {\tmem{Quantum
  Trajectories and Measurements in Continuous Time}}, Vol. 782 of
  {\tmem{Lecture Notes in Physics}} (Springer, Berlin, 2009); A.~Barchielli, in {\tmem{Stochastic Evolution of Quantum States in
  Open Systems and in Measurement Processes}}, edited by L. Di{\'o}si and B.
  Luk{\`a}cs (World Scientific, Singapore, 1994), pp. 1--14

 
\end{thebibliography}
\end{document}